# Interfacial tuning of perpendicular magnetic anisotropy and spin magnetic moment in CoFe/Pd multilayers

D.-T. Ngo,[1] Z. L. Meng,[1] T. Tahmasebi,[1,2] X. Yu,[3] E. Thoeng,[3,4] L. H. Yeo,[4] A. Rusydi,[3,4,a] G. C. Han,[2] and K.-L. Teo[1,b]

[1]Department of Electrical and Computer Engineering, National University of Singapore, 4 Engineering Drive 3, Singapore 117576

[2]Data Storage Institute, A*STAR (Agency for Science Technology and Research), 5 Engineering Drive 1, Singapore 117608

[3]NUSNNI-NanoCore and Singapore Synchrotron Light Source, National University of Singapore, 5 Research Link, Singapore 117603

[4]Department of Physics, National University of Singapore, Singapore 117542

**Abstract**

We report on a strong perpendicular magnetic anisotropy in [CoFe 0.4nm/Pd *t*]$_6$ (*t* = 1.0-2.0 nm) multilayers fabricated by DC sputtering in a ultrahigh vacuum chamber. Saturation magnetization, $M_s$, and uniaxial anisotropy, $K_u$, of the multilayers decrease with increasing the spacing thickness, with a $M_s$ of 155 emu/cc and a $K_u$ of 1.14×10$^5$ J/m$^3$ at a spacing thickness of t = 2 nm. X-ray absorption spectroscopy and X-ray magnetic circular dichroism measurements reveal that spin and orbital magnetic moments of Co and Fe in CoFe film decrease as function of Pd thickness, indicating the major contribution of surface/interfacial magnetism to the magnetic properties of the film.

---
[a] Corresponding author: phyandri@nus.edu.sg
[b] Corresponding author: eleteokl@nus.edu.sg

1. **Introduction**

Perpendicular magnetic anisotropy (PMA) films [1] have been drawn much intention because of their potential applications in magnetic recording technology and spintronics. The PMA films with high saturation magnetization ($M_s$) have been used for perpendicular magnetic recording, bit-patterned media recording technology [2-4]. In spintronics, the PMA films have been exploited in spin-transfer torque (STT) technology, such as STT MRAM [5], nanowire racetrack memory [6], spin logic devices [7], etc. with commitment to create spin-torque devices operating with low energy consumption.

Among the STT applications, the PMA films with unique magnetic properties have been extensively concentrated for spin-torque domain walls (DWs) nanowire devices (non-volatile memory, spin logic devices, etc) [6,7]. One of the most important issues of such devices is to reduce magnetization-switching current density, of which the critical current density is essentially governed by the intrinsic parameters of the films [8]:

$$j_c = \frac{e\alpha M_s^2 \Delta}{g\mu_B P}\gamma_0 \frac{1}{|\beta - \alpha|} \tag{1}$$

Where, $e$ is the electron charge, $g$ is the gyromagnetic ratio, $\mu_B$ is the Bohr magneton, respectively; $\alpha$, $\beta$, $M_s$, $\Delta$ and $P$ are the Gilbert damping factor, non-adiabatic spin transfer parameter, saturation magnetization, domain wall thickness and spin polarization of the film, respectively. In the PMA films, the Bloch-type walls with narrower thickness form instead of Néel-type walls of very large wall thickness which normally exist in the thin films with in-plane anisotropy. The narrow Bloch wall is also favorable for non-adiabaticity that is expected to enhance the wall motion [9], which is crucial for high-speed devices. Therefore, a PMA film with low $M_s$, low damping constant, narrow DW and high spin polarization would be desirable for a spin-torque DW device working with low current density.

In this article, we observe strong PMA on *fcc*-$Co_{70}Fe_{30}$ multilayers with low $M_s$, high uniaxial anisotropy and spin polarization. The low $M_s$ of the multilayers containing high-$M_s$ $Co_{70}Fe_{30}$ as magnetic layers are realized using non-magnetic Pd spacing layers to modify the interfacial magnetic moment proved by X-ray absorption spectroscopy (XAS) and X-ray magnetic circular dichroism (XMCD) measurements.

## 2. Experiments

The CoFe/Pd-based multilayers with the structure of Ta 5.0 nm/Pd 5.0 nm[CoFe 0.4 nm/Pd $t$]$_6$/Ta 2.0 nm ($t$ = 1.0, 1.2, 1.4, 1.6, 1.8 and 2.0 nm, respectively) were grown on thermally oxidized Si substrates by using DC magnetron sputtering in an ultrahigh vacuum chamber of an UHV Deposition Cluster. Composition of the CoFe layers was fixed at $Co_{70}Fe_{30}$ (at.%). The base pressure was better than $2.5 \times 10^{-9}$ Torr whereas Ar working pressure was remained at 1.5 mTorr during deposition. Magnetic properties of the films at room temperature were characterized using an alternating gradient force magnetometer (AGFM) with a maximum magnetic field of 20 kOe.

The X-ray absorption spectroscopy (XAS) and X-ray magnetic circular dichroism (XMCD) at Co $L_{3,2}$ and Fe $L_{3,2}$ edges at room temperature were obtained by recording the sample current (total yield mode) as a function of photon energy at the SINS beam line of the Singapore Synchrotron Light Source (SSLS) [10]. Elliptically polarized light with a degree of circular polarization (DCP) = 80 and an energy resolution of 0.25 eV was employed for the XMCD measurements. To measure the out-of-plane spin and orbital magnetic moments, the light was incident at a normal angle from the sample surface, with its propagation direction along the sample out-of-plane magnetisation direction. An external magnetic field of ±10 kOe was applied to magnetize the sample along the out-of-plane direction. The XCMD were done by changing the direction of applied magnetic field while kept the helicity of the light. Thin films of Co and Fe were used as the reference to normalize the XMCD data. We intentionally measured XMCD with spin polarization along out-of-plane magnetic moments to follow an easy-axis magnetization shown by the magnetometer measurements. Micromagnetic simulation of domain structure was performed using LLG Micromagnetic Simulator$^{TM}$ [11].

## 3. Results and Discussion

Figure 1 shows a series of magnetic-hysteresis (M-H) loops measured on two directions for CoFe/Pd multilayers ($t$ = 1.0÷2.0 nm): parallel (dash curves) and perpendicular (solid-dot curves) to the film plane. The *M-H* loops for the field applied in the film plane exhibit a hard axis behavior with rotation mechanism of the magnetization reversal denoted by S-shape loops. Whilst the loops measured with the perpendicular field show an easy-axis behavior reflected as the highly square *M-H* loops. Smooth *M-H* loops indicate that unique magnetic properties occur over the film as the single-phase behavior due to strong ferromagnetic coupling between the magnetic CoFe layers. These prove the strong alignment of the magnetization of the films perpendicular to the film plane, or the existence of the PMA in the studied multilayers. It should be note that the perpendicular loops are not perfectly 100% squareness which is ascribed to a small in-plane component existing in the films. This in-plane component is attributed to the Néel caps located in the Bloch walls, which will be discussed detailed by looking into the domain structure of the films in coming paragraphs.

Usually, the Co-Fe alloy with a concentration of Fe above 15 at.% is favorable for the formation of body-centered cubic (bcc) crystal structure [12], and the magnetic anisotropy in such a film should appear in-plane. However, the thick Pd seed layer in face-centered cubic (fcc) with a strong <111> texture normal to the film plane was proved in our previous report [13] to induce the <111>-texture fcc structure in the CoFe that supported the out-of-plane anisotropy [12,14]. Hence, the Pd seed layer with <111> texture normal to the film plane is effective to induce the PMA in ultra thin CoFe films [12,14], including CoFe/Pd multilayers [13]. In recent reports, some authors proposed that the interfacial anisotropy of the CoFe/Pd interface, in which the CoFe layer was in <111>-texture fcc, is possibly an important factor governing the PMA of the CoFe/Pd multilayers [15]. Following analysis of magnetic moments by XMCD in this paper allows confirming this prediction, and we prove the minor contribution of the crystal structure of the layers to the PMA of these multilayers.

It is then important to understand the modification of magnetic properties of the CoFe/Pd multilayers by varying Pd spacing layer thickness. Figure 2(a) illustrates the saturation magnetization, $M_s$, of the CoFe/Pd multilayers as a function of Pd spacing layers thickness. Obviously, $M_s$ appears to decrease with the Pd spacing layer thickness due to diluting the magnetic interaction by the non-magnetic Pd layers. With Pd spacing layers of 1.0 nm thick, the $M_s$ value is measured to be 280 emu/cc, whereas this value decreases to 155 emu/cc as the thickness of the spacing layers increases to 2.0 nm. The reduction in $M_s$ is in good agreement with the function of $1/t$ as it has been commonly seen in PMA multilayers [16]. Such $M_s$ values are significantly lower than that in other common Co-based PMA films used for spin-torque domain wall applications, such as Pt/Co multilayers (~900 emu/cc) [17], Co/Ni multilayers (660 emu/cc) [6], perpendicular magnetized CoFeB films (1200 emu/cc) [18,19], and comparable with low-$M_s$ ferrimagnetic TbFeCo film [20,21]. The $M_s$ of the studied CoFe/Pd multilayers is definitely much lower than that of pure fcc-$Co_{70}Fe_{30}$ alloy film (1200 emu/cc) [22]. This reduction is ascribed essentially to the presence of the Pd layers (with weak induced magnetic moments) and also as demonstrated in next section from the XMCD measurements to a decrease of the moments of Fe and Co atoms located at the interfaces.

In contrary to the tendency of the $M_s$, the anisotropy field, which is derived from the M-H loops as $H_k = H_s + 4\pi M_s$ ($H_s$ is the saturation field) [18], increases as a linear function of the Pd layer thickness [see inset of Fig. 2(a)] from 12.6 kOe to 14.8 kOe. From $H_k$ and $M_s$, the uniaxial anisotropy $K_u$ is determined as $K_u = H_k \cdot M_s/2$ [18]. Figure 2(a) shows $K_u$ as a function of the Pd spacing thickness. It is clearly seen that the insertion of Pd space layers reduces the uniaxial anisotropy of the multilayers. As the thickness of the Pd space layers increases from 1.0 nm to 2.0 nm, the $K_u$ decreases from $1.77 \times 10^5$ J/m$^3$ to $1.14 \times 10^5$ J/m$^3$ which is consistent with the $1/t$ law. Such a high $K_u$ is comparable to other common multilayer PMA films used in spintronics (discussed in the previous paragraph). The uniaxial anisotropy allows the domain wall thickness to be estimated using the relation: $\Delta \approx \pi\sqrt{A/K_u}$ with $A$ is the exchange constant. Assuming $A \sim$ 10 pJ/m (typically for Co-based magnetic thin films), the domain wall thickness is about 20-25 nm in scale of that in the Co/Pd multilayers and other Co-based multilayers [23]. Additionally, the exchange interaction length can be deduced from the $M_s$ by using the relation $l_{ex} = \sqrt{2A/\mu_0 M_s^2}$. This value is in range of 7-12 nm, which is about the overall thickness of the multilayers, supporting the thin magnetic layers in the multilayers to ferromagnetically couple to each other.

To understand the magnetic anisotropy in our samples, micromagnetic simulations of domain structure for out-of-plane and in-plane induction components were performed using LLG commercial simulator [11]. Figure 3 depicts simulated domain pattern in the multilayer with the Pd thickness of 1.0 nm at demagnetized state as a typical picture (the simulated domain patterns of the other samples look similar to this picture – data not shown). A stripe-like domain structure, which looks similar to that of the PMA FePd film reported previously [24] is observed in all the multilayer samples with a very well-defined period of about 400 nm, and mostly unchanged with the Pd spacing thickness. This is also the size of the domains in some PMA multilayers based on

Co [19,25]. For the stripe domains, it is important to take into account the quality factor which is given by [25]:

$$Q = \frac{2K_u}{\mu_0 M_s^2} \qquad (2)$$

In our studied multilayers, the $Q$ factor ranges from 3.5 to 7.6, indicating that the anisotropy energy term is dominant to form a sharply defined domain state as seen in Fig. 3(a) with the Bloch-type domain walls extending right up the surface of the films with a weak Néel caps in similar size of the Bloch walls as described in previous reports [24,26]. The simulated results shown in Fig. 3 somehow affirm this suggestion. Namely, perfectly perpendicular magnetization in the domains is obviously visible via the blue-red color whilst the Néel caps are visualized as the in-plane components located at the domain walls [Fig. 3(b)]. These in-plane Néel caps are supposed to be the contribution to the less-squareness in the perpendicular loops as seen in Fig. 1. In the application points of view, if the films are used in domain wall devices, the Néel caps would help the walls to nucleate easier.

The final important aspect is to discuss about the electronic structure and spin polarization of the films revealed using XAS and XMCD. Fig. 4(a,b) shows the XAS and XMCD at Fe $L_{3,2}$ (i.e. Fe $2p$ → Fe $3d$ transitions) and Co $L_{3,2}$ edges (i.e. Co $2p$ → Co $3d$ transitions). Thus, because of dipole selection rule, these transitions are element specific and extremely sensitive to electronic structure and spin polarization at the Co $3d$ and the Fe $3d$ bands. Due to strong core hole spin-orbit coupling, XAS at Co $L_{3,2}$ edges show main two peaks, at ~778 eV for $L_3$ and at ~794 eV for $L_2$, while XAS at Fe $L_{3,2}$ edges shows main two strong peaks, at ~708 eV for $L_3$ and ~721 eV for $L_2$ in the CoFe layers. In Fig. 4(a), we present XAS (using circular polarized light with different direction of magnetic field with respect to the normal surface of sample) and XMCD on CoFe layers of the multilayer film with a Pd spacing thickness of 1.0 nm as a typical example. A magnetic field of ±10 kOe is applied out-of-plane direction and the XMCD signal is deduced by subtracting two XAS signals at the opposite magnetic field directions, namely positive and negative fields. We have fabricated Co and Fe films (~200 nm thick) and used them as the reference samples to calibrate and compare XAS and XMCD signals [Figs. 4(c,d)]. In the XMCD signal, we observe strong ferromagnetic properties in the films, which can be referred as two well-defined peaks at both edges, $L_3$ and $L_2$ of Co and Fe. This is strong evidence that the intrinsic ferromagnetism in our CoFe/Pd multilayers results from the Co $3d$ and Fe $3d$ states, essentially. The enhancement of the normalized peak in the XMCD signal of the CoFe/Pd multilayer films comparing to those of Co and Fe reference signals [Figs. 4(c,d)] indicates the strong ferromagnetic interaction between Co-Fe ions in the CoFe lattice which is well-known as the origin of the high magnetic moment in the Co-Fe alloy system [27].

One of great advantages of XMCD data is its capability to reveal the spin and orbital magnetic moments [28]. By applying the X-ray MCD sum rule [29,30], we have estimated spin magnetic moment ($\mu_S$) and orbital magnetic moment ($\mu_L$) based on the following equations:

$$\mu_S = -\frac{1}{\cos\theta \times CPD} \times \left(\frac{6\int_{L_3}(\mu_+ - \mu_-)d\omega - 4\int_{L_3+L_2}(\mu_+ - \mu_-)d\omega}{\int_{L_3+L_2}(\mu_+ + \mu_-)d\omega}\right) \times (10 - n_{3d})\left(1 + \frac{7\langle T_z \rangle}{2\langle S_z \rangle}\right) \tag{3}$$

$$\mu_L = -\frac{1}{\cos\theta \times CPD} \times \left(\frac{4\int_{L_3+L_2}(\mu_+ - \mu_-)d\omega}{3\int_{L_3+L_2}(\mu_+ + \mu_-)d\omega}\right) \times (10 - n_{3d}) \tag{4}$$

where $n_{3d}$ is the 3d electron occupation number, $\langle T_z \rangle$ is the expectation value of magnetic dipole operator, $\langle S \rangle$ is equal to half of the $m_{spin}$ in Hartree atomic units, $\theta$ is photon incident angle which is 0°, and CPD is circular polarization degree which is 0.8. Based on band structure calculations, the $\langle T_z \rangle/\langle S_z \rangle$ is negligible. From the elementary magnetic moments of Co and Fe atoms, magnetic moments of the $Co_{70}Fe_{30}$ composition in the CoFe/Pd multilayers are calculated by combining those of Co and Fe. The results are shown in Figures 4(e-g).

The spin magnetic moment observed in the XMCD signals [Figs. 4(e,g)] reveals that spin polarization (*P*), which is associated to the spin magnetic moment ($\mu_s$) [31] ($P = \left(N^{\uparrow} - N^{\downarrow}\right)\mu_s$), is evidenced. As a result, by normalizing the spin magnetic moment of the CoFe/Pd multilayer films to the reference Co and Fe films, the spin polarization in the CoFe/Pd multilayer films can be estimated. It is assumed that the spin polarization of Fe is ~43-45% [22,27]. Then, the spin polarization in the studied CoFe/Pd-based multilayers can be qualitatively evaluated to be ~45-60%, and better than the spin polarization of the Co-based multilayers (e.g. 56% in Pt/Co multilayer) [32]. It should make a note that the CoFe-based multilayer obtained in this study can be competitive with the well-known Co-based multilayers for the spin-transfer torque applications because of their controllably low $M_s$, high $K_u$, narrow Bloch wall and high spin polarization. From the technical point of view, using of alloy layers (in this case $Co_{70}Fe_{30}$ ally layers) would bring benefit in stabilization of device structure and properties. Commonly, layers of the multilayers are in few Angstrom thick and the PMA would be logically threatened by heat released from the electrical current due to the diffusion between the layers. Close packed structure of the alloy magnetic layers would help to prevent the diffusion to enhance the thermal stability of the devices.

Interestingly, Figs. 4(e,g) show that spin magnetic moment and total magnetic moment apparently decrease by varying the Pd spacing thickness. This is in good agreement with our AGFM result [Fig. 2(a,c)]. Because the thickness of the CoFe magnetic layers is fixed, such decreases can be assigned to the contribution of the interfacial magnetic moment of the CoFe/Pd interfaces, which varies with Pd thickness. This is an evidence to prove that the interfacial anisotropy significantly affects the magnetic anisotropy of the films.

## 4. Conclusions

In conclusion, the $Co_{70}Fe_{30}$/Pd multilayers with strong perpendicular magnetic anisotropy and modifiable magnetic properties have been fabricated. Saturation magnetization as low as 155 emu/cc and narrow Bloch-wall type as thin as 20-25 nm are obtained by increasing the Pd layer thickness to 2.0 nm to reduce the interlayer exchange coupling. Using XAS and XMCD measurements, spin polarization is observed and estimated (~43-60%) to be higher in the very thin CoFe layers than those in the thicker Co and Fe films. Furthermore, the XAS and XMCD spectra reveal the modification of electronic structure of CoFe/Pd interfacial magnetic moments by changing the Pd spacing layer, contributing to the magnetic properties of the films. These advantages indicated that our studied CoFe/Pd multilayers are of superior characteristics for spin-torque applications.

**ACKNOWLEDGEMENTS**

This work was completed by a financial support from the A*STAR (SERC) Public Sector Funding (Grant No. 092 151 0087). We also acknowledge MOE-AcRF-Tier-2 and CRP Awards No. NRF-CRP 8-2011-06 and NRF-CRP 8-2009-024 for the works performed at Singapore Synchrotron Light Source. D.-T. Ngo would like to thank Prof. Michael R. Scheinfein (Arizona State University in Tempe, Arizona) for his kind support in LLG simulation.

**Figure captions**

FIG. 1. Magnetic hysteresis (M-H) loops of the CoFe/Pd multilayers with different Pd spacing thickness. The loops with circle dots were measured out-of-plane of the films, and the dash plots denoted the in-plane loops.

FIG. 2. Linear dependence of the saturation magnetization, $M_s$ (a) and the uniaxial anisotropy, $K_u$ (b) on the inversed thickness ($1/t$) of Pd spacing layer. The inset shows anisotropy field, $H_k$, as a function of Pd spacing thickness.

FIG. 3. Simulated domain pattern in the CoFe/Pd multilayer with a spacing thickness of 1.0 nm: (a) out-of-plane induction component and (b) in-plane induction components. The images size is 10 µm.

FIG. 4. (a,b) X-ray absorption (XAS) and X-ray magnetic circular dichroism (XMCD) spectra of the Fe and Co atoms in the CoFe/Pd multilayer with a spacing thickness of 1.0 nm. (c,d) XAS and XMCD spectra of the Fe and Co atoms in thick reference Fe and Co films. (e,f) Particular elementary spin magnetic moment ($\mu_s$), orbital magnetic moment ($\mu_L$) of Fe and Co atoms in the CoFe/Pd multilayers derived from XMCD spectra. (g) Spin magnetic moment, orbital moment and total magnetic moment of $Fe_{70}Co_{30}$ composition in the CoFe/Pd multilayers.

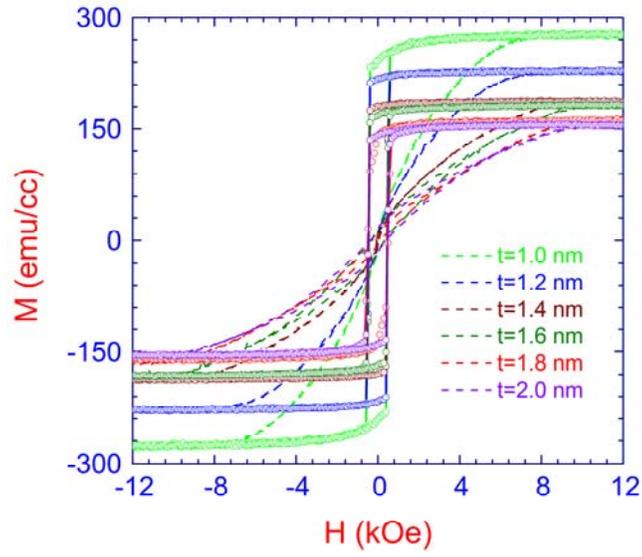

**FIG. 1.**

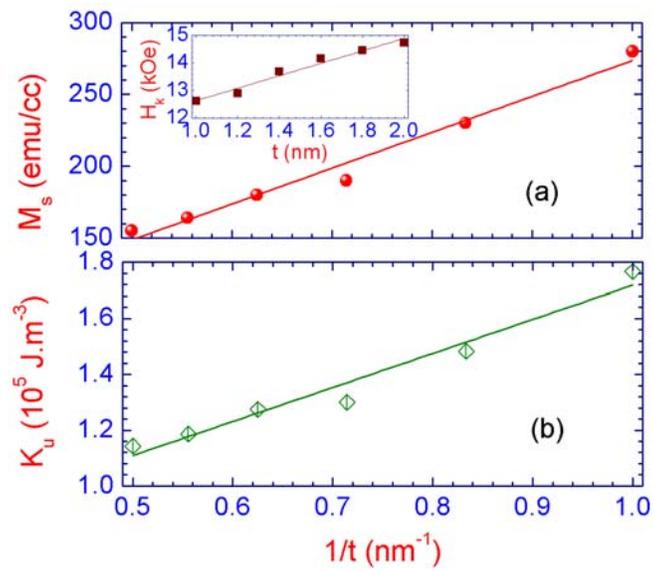

**FIG. 2.**

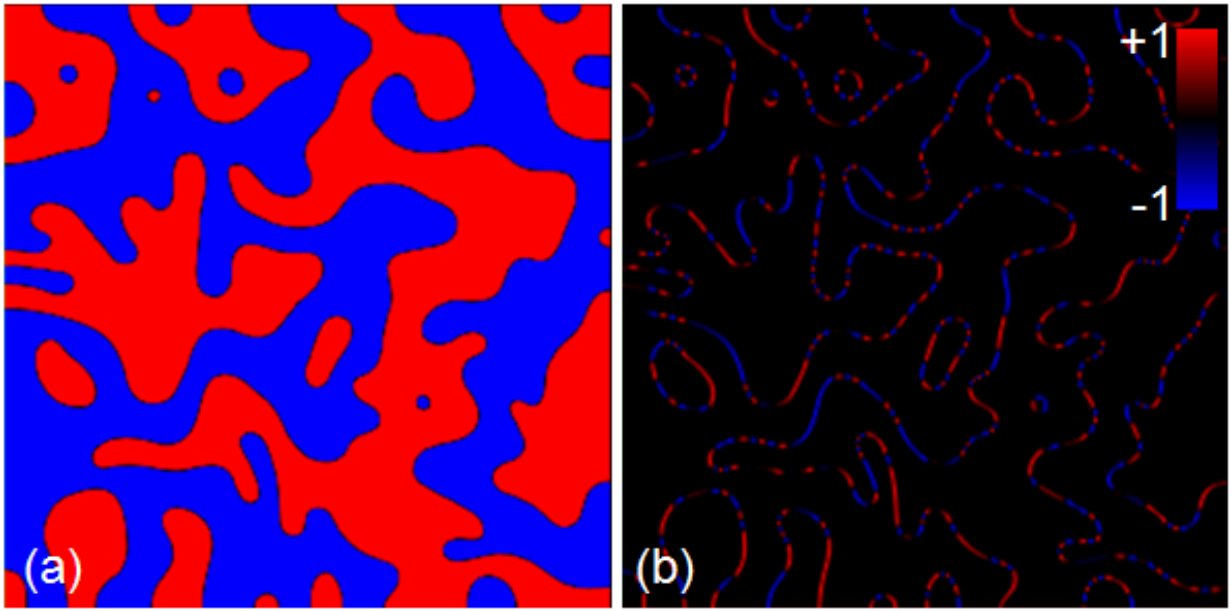

**FIG. 3.**

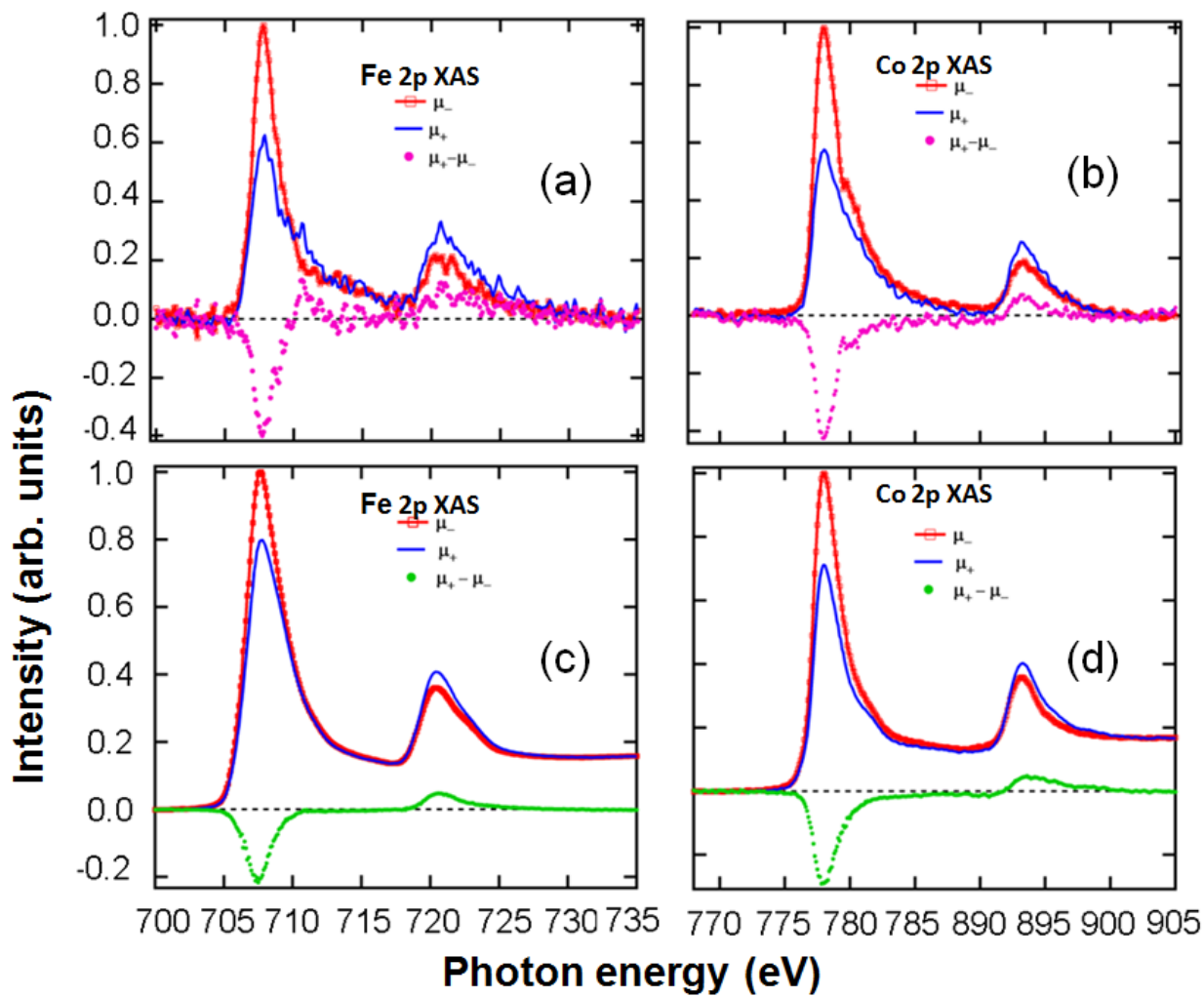
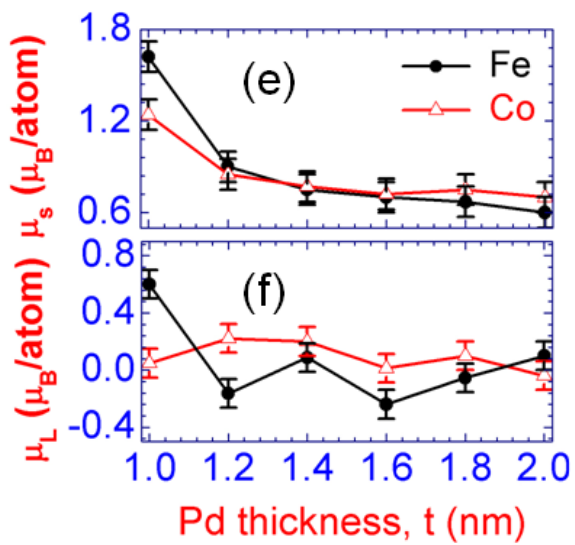
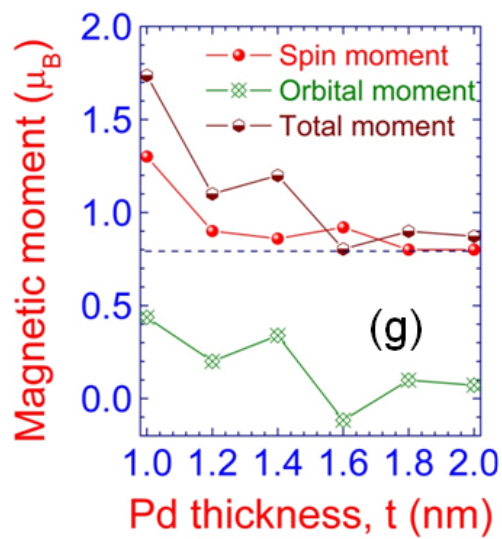

**FIG. 4.**